\documentclass[letterpaper, 10 pt, conference]{ieeeconf}  

\IEEEoverridecommandlockouts                              

\usepackage{graphics} 
\usepackage{epsfig} 
\usepackage{amsmath} 

\usepackage{xcolor}
\usepackage{color,soul}
\usepackage{lipsum} 
\usepackage{hyperref}
\usepackage{multirow} 
\usepackage{soul}

\title{\LARGE \bf
Efficient Face Detection with Audio-Based Region Proposals\\for Human-Robot Interactions
}

\author{William Aris and François Grondin$^{1}$
\thanks{*This work was supported in part by the Natural Sciences and Engineering Research Council of Canada - Collaborative Research and Training Experience program.}
\thanks{$^{1}$W. Aris and F. Grondin are with the Department of Electrical Engineering and Computer Engineering, Université de Sherbrooke, Québec (Canada) J1K 0A5
        {\tt\small \{William.Aris, Francois.Grondin2\}@USherbrooke.ca}, }%
}

\begin{document}

\maketitle
\thispagestyle{empty}
\pagestyle{empty}

\begin{abstract}


Efficient face detection is critical to provide natural human-robot interactions. However, computer vision tends to involve a large computational load due to the amount of data (i.e. pixels) that needs to be processed in a short amount of time. This is undesirable on robotics platforms where multiple processes need to run in parallel and where the processing power is limited by portability constraints. Existing solutions often involve reducing image quality which can negatively impact processing. The literature also reports methods to generate regions of interest in images from pixel data. Although it is a promising idea, these methods often involve heavy vision algorithms. In this paper, we evaluate how audio can be used to generate regions of interest in optical images to reduce the number of pixels to process with computer vision. Thereby, we propose a unique attention mechanism to localize a speech source and evaluate its impact on an existing face detection algorithm. Our results show that the attention mechanism reduces the computational load and offers an interesting trade-off between speed and accuracy. The proposed pipeline is flexible and can be easily adapted to other applications such as robot surveillance, video conferences or smart glasses. 

\end{abstract}

\section{INTRODUCTION}
In human-robot interaction, it is crucial for the robot to focus its attention and engage as much resources as possible quickly on the active user to provide a natural interaction. Face detections and various computer vision techniques are often at the heart of this process \cite{andrist2014robots}.
However, due to the large amount of data to process in a short amount of time, i.e. thousands of pixels per frame, it often involves a large computational load.
For example, many methods such as convolutional neural networks (CNN) \cite{cun_handwritten_1990}, edge detection techniques \cite{canny_computational_1986} and the Viola-Jones face detector \cite{viola_rapid_2001-1} rely on the convolution operation.
Each convolution involves sliding a kernel along the image leading to a time complexity of approximately $\mathcal{O}(K^2HW)$ where $K$ is the kernel size, $H$ is the image height and $W$ is the image width.
While this is for one single convolution, most algorithms make use of multiple kernels, which further increases the number of computations.
This can become a problem when working with high resolution images and/or low-cost embedded systems.

One way to tackle this challenge is to decrease the image resolution by using techniques such as nearest neighbor or bilinear interpolations.
However, this often leads to a degradation of the image quality, which can negatively impact the performance of certain algorithms \cite{aqqa_understanding_2019}.
Another strategy consists in defining region proposals in the image and processing these regions with more expensive techniques. 
Selective search \cite{uijlings_selective_2013} and region proposal networks (RPN) \cite{ren_faster_2015, kong_hypernet_2016} are probably the most popular methods used for this purpose. 
Selective search first segments the image based on pixel intensity before grouping similar segments iteratively. Although this method has a high recall, it usually requires several seconds to perform computations for a low-resolution image \cite{hosang_how_2014, kong_hypernet_2016}, which makes real-time processing impossible. RPNs are often implemented as small sub-CNNs that take as input an image and output regions of interest (ROI) where the targeted objects could be. They tend to be faster than selective search and can run on low-cost hardware. However, this strategy still involves processing the entire image with convolution operations, which is not scalable. It's also worth mentioning that a third method, edge boxes \cite{zitnick_edge_2014}, is faster than selective search but, according to the results reported in \cite{hosang_how_2014}, fails to run in real-time.

Moreover, traditional region proposal techniques can determine \textit{where} in an image a robot should engage its resources, but they ignore \textit{when} it is worth doing so. 
In the case of human-robot interactions, it is often unnecessary for robots to continuously monitor the visual scene and look for faces when nobody is talking to them.
Context awareness is key to share computing power amongst multiple processes and extend battery life. 

This work demonstrates how sound can be used effectively to make region proposals around a speaker's face and reduce the computational load in the context of human-robot interactions. Audio signals have the advantage of being represented by fewer data points than optical images while providing information redundancy at the semantic and spatial levels.  For example, when a student asks a question in a classroom, the teacher can identify who is talking either with visual information, sound cues or both.
We limit ourselves to the case of a single speaker in various noisy environments for this work. However, the proposed pipeline stays relevant in many scenarios and because of its modularity, it can easily benefits from incremental research and eventually be applicable to situations with multiple speakers talking at the same time. In addition, the pipeline can easily be adapted to other applications such as robot surveillance, video-conferences or smart glasses for the visually impaired.

The literature already reports techniques that use audio jointly with visual data.
Both modalities are usually processed in parallel and late fusion increases detection robustness \cite{zhang_boosting-based_2008, lacheze_audiovideo_2009, hutchison_audio-visual_2010}. However, late fusion usually increases the computational load. 
To avoid redundancy, the alternative is to use the two modalities in cascade to complement one another. 
For example, an omnidirectional camera can be used to localize faces, whose positions are fed to a beamformer to find active speakers \cite{kapralos_audiovisual_2003}. 
In \cite{yoshimi_multimodal_2002}, the authors go the other way around and use the direction of arrival of sound to select a static camera and perform face detection. 
Finally, in \cite{barnard_audio-visual_2013} an audio tracker is used to limit the search of a face detection algorithm.
This work introduces two novel components: 1) we propose a method to determine when the robot should engage more resources in generating ROIs and 2) we introduce a time-frequency attention mechanism to localize the target speech source, and ignore undesirable interferences. 

This paper is organized as follows. Section II gives an overview of the system and explains the different design choices, Section III explains the implementation details of each module, Section IV presents the experimental results obtained and Section V concludes the paper and proposes avenues for future work.

\section{VOICE-TARGETED REGION PROPOSALS MECHANISM}

\begin{figure}[!htbp]
    \centering
    \includegraphics[width=0.85\columnwidth]{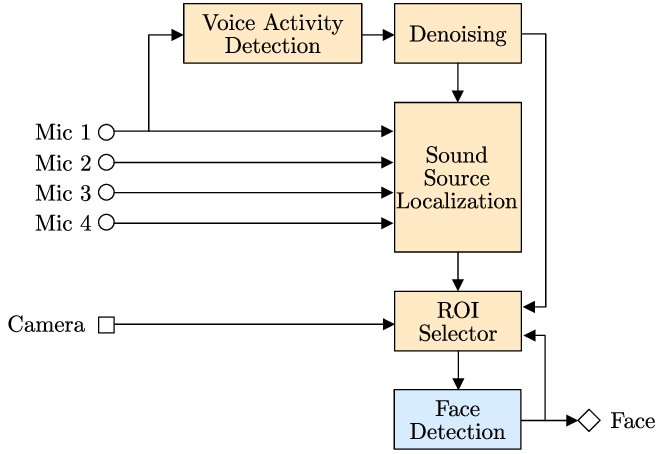}
    \caption{Overview of the proposed pipeline.}
    \label{fig:pipeline}
\end{figure}

Figure \ref{fig:pipeline} shows the global architecture of the system.

The pipeline first includes a simple voice activity detection (VAD) module to act as a gate-keeper. The module is here to add a layer of efficiency and indicate when the context is right to use the rest of the pipeline. If no voice is detected, it can be assumed that nobody wants to interact with the robot and the system can go in idle mode.
Once voice is detected, audio is sent to the denoising module. Audio is often a mixture of the desired signal and noise which can degrade the localization performance. When the sound-to-noise (SNR) is low, it becomes essential to denoise the raw signal.

\begin{figure}[!htbp]
    \centering
    \includegraphics[width=0.9\columnwidth]{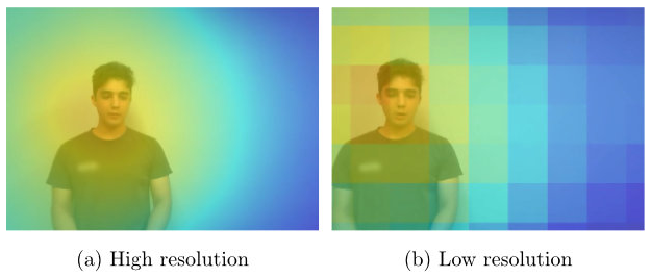}
    \caption{Acoustic image example with with resolutions of (a) $640 \times 480$ regions and (b) $9 \times 7$ regions.}
    \label{fig:acoustic-images}
    \vspace{-5pt}
\end{figure}

The sound source localization (SSL) module uses the denoised signal to localize voices. Spatial information from audio can easily be derived using a microphone array and by exploiting the time difference of arrival (TDOA) of each microphone pair. The literature reports many SSL techniques \cite{knapp_generalized_1976, schmidt_multiple_1986, dibiase_high-accuracy_2000}, each having pros and cons. The steered-beamformer based approach is particularly interesting in this case because the search space is well constrained by the image plane and the number of directions to look at can be parametrized to fit the requirements of specific situations (i.e., the resolution of the localization can be adjusted to save computation time without affecting the resolution of the optical image). By projecting the result of the steered-beamformer on an image plane, it is possible to generate acoustic images similar to figure \ref{fig:acoustic-images}. An acoustic image can be seen as a heatmap showing where sound power lies with respect to an optical image. The resolution of the acoustic images can be adjusted without affecting the resolution of the optical images.

Finally, the ROI selector uses information provided by the denoising module and the previous face detection result to determine an appropriate size for the ROI. It then uses the SSL module to extract the region from the image at the right location. An interesting by product of this mechanism is that it makes it possible to detect if a speaker is outside the field of view of the camera and to reorient the camera toward this person if necessary.

\section{SYSTEM IMPLEMENTATION}
\subsection{Hardware}
\begin{figure}[!htbp]
    \centering
    \includegraphics[width=0.6\columnwidth]{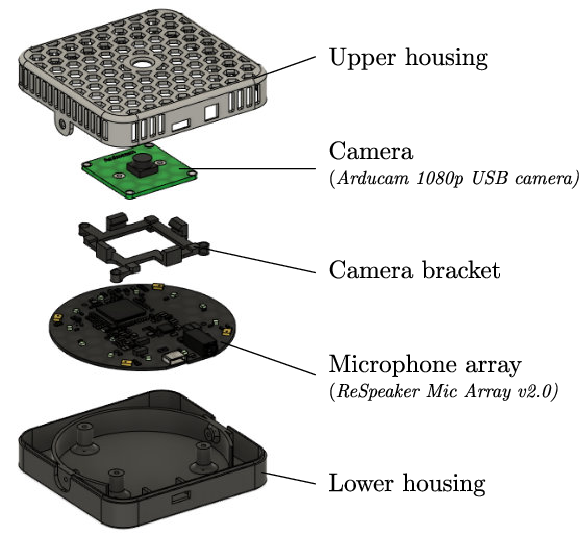}
    \caption{Acoustic camera developed for the project.}
    \label{fig:acoustic-camera}
\end{figure}

We design the low-cost acoustic camera shown in figure \ref{fig:acoustic-camera} for the project. The device is mainly based on an Arducam 1080p USB camera and a 4-channels ReSpeaker mic array. The housing is entirely 3D printed.

To generate representative datasets when training neural networks, we measure 1000 multichannel room impulse responses (RIRs) with it. 
The RIRs are obtained using the exponential sine sweep (ESS) method proposed by Farina in \cite{farina_simultaneous_2000} as it removes some distortions and offers a good SNR.
When measuring the RIRs, a Bose SoundLink Flex loudspeaker generates 10~seconds ESSs going from 20~Hz to 8000~Hz.
Recording is performed at a sample rate of 16~kHz and stops 3~seconds after the end of the signal to capture the late reverb.
A total of 10 different rooms (100 samples per room) with different acoustic properties are sampled. 
The loudspeaker is moved between each measurement at random positions and the acoustic camera is moved between 3 to 4 times in each room.

This low-cost acoustic camera becomes a convenient tool for research in multimodal perception. 
A public repository contains the CAD files to build a camera, the measured RIRs and the configurable python scripts to capture new RIRs if necessary \footnote{\url{https://github.com/introlab/echolense}}.

\subsection{Voice activity detection}
\begin{figure}[!htbp]
    \centering
    \includegraphics[width=0.7\columnwidth]{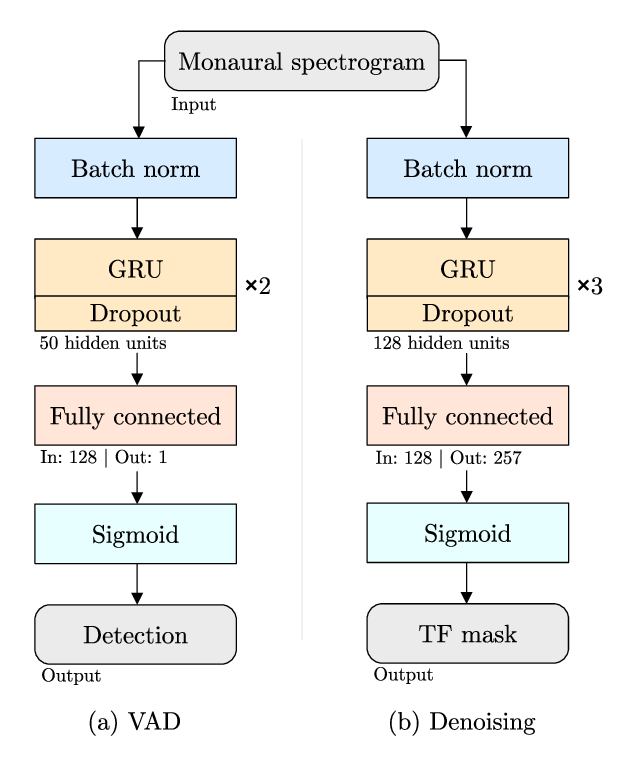}
    \caption{Architecture for the VAD module (a) and the denoising module (b). The VAD network has 61,703 parameters while the denoising network has 379,907 parameters.}
    \label{fig:neural-networks}
\end{figure}

At its core, the VAD module uses a small recurrent neural network. As shown in figure \ref{fig:neural-networks}, the network consists of two gated recurrent units (GRU) layers so that it can extract patterns over time, a linear layer to synthesize the information from the latent space and a sigmoid to scale the output between 0 and 1.

\vspace{-5pt}

\begin{figure}[!htbp]
    \centering
    \includegraphics[width=0.9\columnwidth]{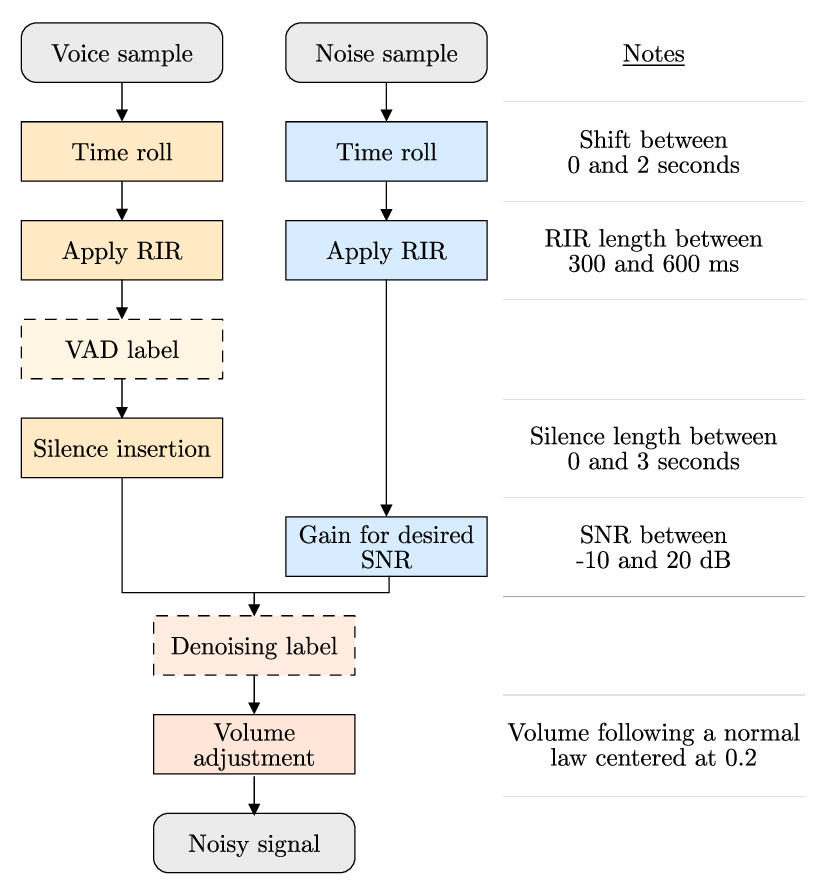}
    \caption{Data augmentation to generate training samples for VAD and denoising modules -- 10\% voice only, 15\% noise only, 40\% voice + FSD50K noise, 30\% voice + MusicNet noise and 5\% voice + white noise.}
    \label{fig:augmentation-pipeline}
\end{figure}

The VAD neural network is trained on a 400 hours audio dataset with 280 hours used for training and 120 hours used for validation. The samples are obtained through the data augmentation pipeline presented in figure \ref{fig:augmentation-pipeline}. 
Most of the samples contain vocals from LibriSpeech \cite{panayotov_librispeech_2015}, mixed with noise from FSD50K \cite{fonseca_fsd50k_2022}, classical music from MusicNet \cite{thickstun_musicnet_2016} or white noise artificially generated. 
The rest of the dataset is made of samples containing either only voice or only noise. 
As music tends to generate more false alarms when compared to the other noise types, samples are generated more often with music as background noise than with other noise types.

\vspace{-5pt}

During data augmentation, each voice sample and noise sample is randomly rolled (circular time shift) to introduce diversity in time alignment.
Silence is inserted in every 1 out of 2 voice samples so that the model learns that speech can be discontinuous. 
The samples are convolved with random measured RIRs. 
Finally, the SNR is chosen in the interval $[-10, +20]$ dB, and the signals are added together.
The volume is adjusted following a Gaussian distribution, which parameters were obtained from preliminary recordings made with the camera.

Equations \ref{eq:vad_noisy}-\ref{eq:vad_target} show the process to obtain the training target (VAD label) from the voice samples. The energy of each STFT time frame is first compared to the first quartile of the energy of all frames . If the value for a given frame is above the first quartile, it is assumed that this frame contains voice. This works because a good portion of the time frames contains voice. That is also why the target is generated before the silence insertion (the label is adjusted to match the silence afterward). A moving average over 10 time frames is applied to the label to make it robust to short breaks. Finally, the target is binarized with a threshold set at 0.5:

\begin{equation} \label{eq:vad_noisy}
VAD_N[l] = \begin{cases}
    1, & E[t] \geq Q_1 \\
    0, & E[t] < Q_1
    \end{cases},
\end{equation}
\begin{equation} \label{eq:vad_desensitized}
VAD_D[t] = \frac{1}{10} \sum_{j=t-9}^{t}VAD_N[j],
\end{equation}
\begin{equation} \label{eq:vad_target}
VAD[t] = \begin{cases}
    1, & VAD_D[t] \geq 0.5 \\
    0, & VAD_D[t] < 0.5
    \end{cases} ,
\end{equation}
where $E[t]$ is the energy of the time frame $t$, $Q_1$ is the first quartile of the energy of all the frames, $VAD_N$ is the noisy label, $VAD_D$ is the desensitized label and $VAD$ is the training label.

\begin{figure}[!htbp]
    \centering
    \includegraphics[width=\columnwidth]{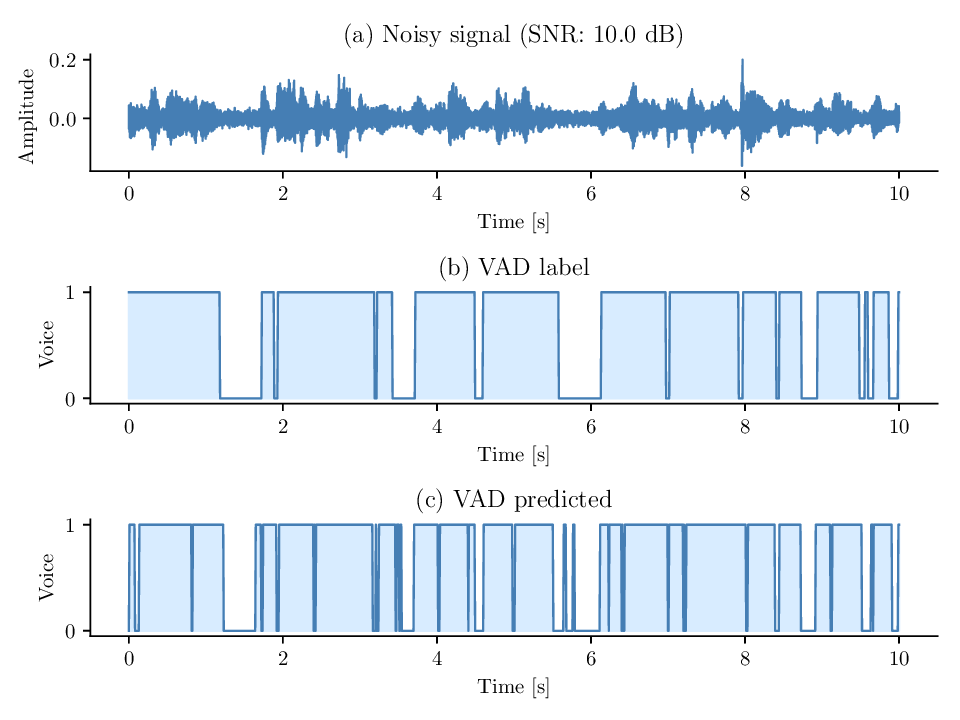}
    \caption{VAD example with a SNR of 10 dB and a threshold of 0.3.}
    \label{fig:example-vad}
    \vspace{-10pt}
\end{figure}

A binary cross-entropy loss function is used and the model is trained over 100~epochs with the Adam optimizer, a batch size of 64 and a learning rate of 0.001.
The final model obtained an accuracy score of 93.4\% on the validation dataset. Figure \ref{fig:example-vad} shows an example of VAD predicted by the model on data recorded with the acoustic camera.

\subsection{Denoising}

As shown in figure \ref{fig:neural-networks}, the denoising module uses a neural network similar to the VAD module. The main difference is that the GRU layers contain more hidden units since the task is more complex: the model needs to generate a time-frequency (TF) mask from a given spectrogram to attenuate the noise bins and preserve the voice using (\ref{eq:denoising}). 
Here $\hat{S_s}$ and $X_s$ are the spectra of the denoised signal and noisy signal, respectively, and $k$ is the frequency bin index, $t$ is the frame index and $M$ is the time-frequency mask:
\begin{equation} \label{eq:denoising}
\hat{S_s}[t, k] = M[t, k] X_s[t, k].
\end{equation}

As opposed to the speech enhancement systems, the predicted mask can introduce distortions in the reconstructed speech signal, as long as it produces accurate localization results. 
After experimenting with the different types of masks listed in \cite{erdogan_phase-sensitive_2015}, we found out that the combination of (\ref{eq:soft-mask}) and (\ref{eq:mask}) provides the best  localization results for this application:
\begin{equation} \label{eq:soft-mask}
M_{soft}[t, k] = \frac{|S_s[t, k]|^{2}}{|X_s[t, k]|^{2}}\cos(\angle{S_s[t, k]} - \angle{X_s[t, k]}),
\end{equation}
\begin{equation} \label{eq:mask}
M[t, k] = \begin{cases}
    1, & M_{soft}[t, k] \geq 0.5 \\
    0, & M_{soft}[t, k] < 0.5
    \end{cases},
\end{equation}
where $S_s$ is the spectrum of the voice signal, $\angle$ stands the phase of a complex number, $M_{soft}$ is a phase sensitive mask and $M$ is a binarized version of it and the training target.
The choice of (\ref{eq:soft-mask}) is also confirmed in \cite{wang_robust_2019} for a similar task.

\begin{figure}[!htbp]
    \centering
    \includegraphics[width=\columnwidth]{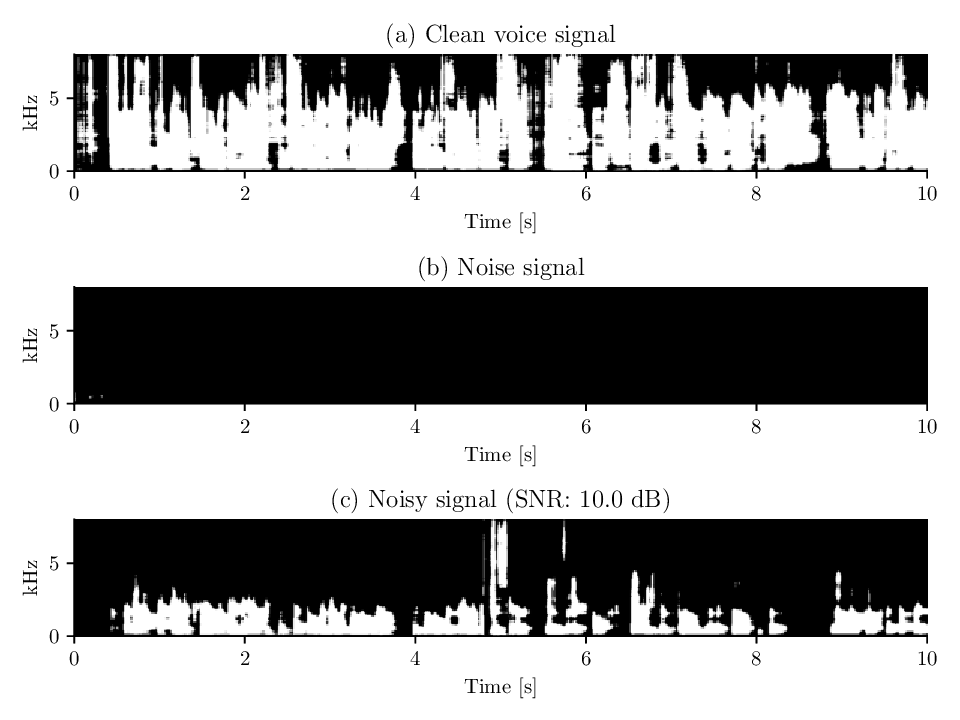}
    \caption{Generated masks with the denoising module at an SNR of 10 dB and a threshold set at 0.7. Black stands for 0 and white for 1.}
    \label{fig:example-denoising}
    \vspace{-10pt}
\end{figure}

The model is trained with the same parameters as the VAD module.
It produces an accuracy score of 85.3\% on the validation dataset. Figure \ref{fig:example-denoising} shows examples of masks estimated by the network. 

\subsection{Acoustic images}

The acoustic image is generated using the SVD-PHAT method \cite{grondin2020audio}.
In this work, the resolution of the acoustic image is set to be $9 \times 7$ regions (as shown in Fig. \ref{fig:acoustic-images}).
The average cross-spectrum with phase transform between each pair of microphones denoted by indexes $i$ and $j$ is obtained recursively using the denoising mask $M[t,k]$:

\begin{equation}
    X_{i,j}[t,k] = (1-\alpha) X_{i,j}[t-1,k] + \alpha M[t,k]\hat{X}_{i,j}[t,k],
\end{equation}
where
\begin{equation}
    \hat{X}_{i,j}[t,k] = \frac{X_i[t,k]X_j[t,k]^*}{|X_i[t,k]||X_j[t,k]|},
\end{equation}
and $\alpha$ is the adaptive rate, and $\{\dots\}^*$ stands for the complex conjugate operator.
The SVD-PHAT method generates efficiently the acoustic image for each region using a low-rank projection matrix and a supervector that holds the cross-spectra for all pairs and frequency bins \cite{grondin2019svd}.
The region with the maximum acoustic energy corresponds to the sound source position.

\subsection{Regions of interest selector}

The selector first updates the size of the ROI by computing a \textit{speech dominance score} (SDS) from the soft values of the denoising mask:

\begin{equation}
    D_s = \frac{1}{T K} \sum_{t=1}^{T} \sum_{k=1}^{K} M[t,k]
\end{equation}

The SDS acts as an indicator to estimate how much of the spectrogram is dominated by speech. Here $D_s$ stands for the SDS, $T$ for the number of time frames and $K$ for the number of frequency bins. The result is then compared to a threshold set to 0.3 empirically. If the SDS is superior to the threshold and a face was detected in the previous optical frame, the ROI size is reduced by a factor of $\beta = 0.1$. If the previous conditions are not met, the size is enlarged. We found empirically the optimal ROI dimensions between 35\% and 65\% of the original image size.

\section{EXPERIMENTAL RESULTS}

\subsection{Evaluation dataset}




We validated the proposed approach in realistic conditions using a dataset of 500 videos generated from recordings made with the acoustic camera. The recordings include 6 individuals (3 males / 3 females) reading prompts in French at different positions in front of the acoustic camera, and 3 noise sources: jiggling keys, a hairdryer blowing and classical music playing in a loudspeaker. This gives a total of 117 voice recordings and 221 noise recordings divided in 3 rooms. The images were captured at a rate of approximately 15 frames per second, with a resolution of  $640 \times 480$~pixels.

The audio of each video in the dataset was generated by randomly selecting a voice sample and a noise sample from the same room, and by mixing them together at a desired SNR. The volume is set to 0.2. The images of the videos come from the voice recordings and contain only the face of the individuals speaking.

\subsection{Modules validation}

To evaluate the VAD, we used equations \ref{eq:vad_noisy}-\ref{eq:vad_target} on the dataset to determine the targets and compare them to the output of the module. At a time-frame level, the dataset is unbalanced as more frames contain voice. Therefore, to get a representative overview of the model, we use the accuracy, the sensitivity and the specificity as metrics. We also measure the specificity on the noise recordings to determine if the model had more difficulties with certain types of noise. Table \ref{tab:vad-metrics} shows the results.

\begin{table}[htbp]
\renewcommand{\arraystretch}{1.3}
\caption{VAD module performance with threshold at 0.3}
\vspace{-10pt}
\begin{center}
\begin{tabular}{|c|c|c|c|}
\hline

Types of samples & Accuracy & Sensitivity & Specificity \\
\hline
\hline

Mix @ SNR 20~dB&0.8493&0.8708&0.7641\\

Mix @ SNR 10~dB&0.8385&0.8651&0.7333\\

Mix @ SNR 0~dB&0.7858&0.8184&0.6567\\

Jiggling keys only&-&-&0.9425\\

Hairdryer only&-&-&0.6948\\

Music only&-&-&0.5018\\
\hline


\end{tabular}
\vspace{-10pt}
\label{tab:vad-metrics}
\end{center}
\end{table}


According to these results, the VAD module seems biased toward detecting voices and generating false positives. This behavior is more desirable than the opposite (generating false negatives) as it forces the system to generate ROIs in case of doubt. This explains why the threshold is set to a low value (0.3 in this case). This bias could also be due to the training dataset being imbalanced at a time-frame level, but further investigations are needed to confirm this with certainty.

The performance of the denoising module for localization purpose is evaluated by measuring the average distance (in pixels) between the estimated direction of arrival (DoA) of the clean voice and the estimated DoAs of the noisy and denoised signals for each video frame.
A paired t-test is also performed to ensure the validity of the observed difference. 
Table \ref{tab:denoising-metrics} shows the results. 
These results confirm that the denoising module is able to reduce the distance by a significant amount even in low SNR. 

\begin{table}[htbp]
\renewcommand{\arraystretch}{1.3}
\caption{Denoising module performance with threshold at 0.7}
\vspace{-10pt}
\begin{center}
\begin{tabular}{|c|c|c|c|c|c|}
\hline
\multirow{2}{*}{SNR} & \multicolumn{2}{|c|}{Distance from clean signal DoA} & \multirow{2}{*}{t-test} & \multirow{2}{*}{$\boldsymbol{p < 0.05}$}\\
\cline{2-3} 

 & Noisy signal & Denoised signal & &\\
\hline
\hline

20~dB&75.10~px&38.94~px&11.44&Yes\\

15~dB&122.79~px&57.73~px&15.88&Yes\\

10~dB&165.66~px&84.50~px&18.05&Yes\\

5~dB&201.08~px&117.14~px&18.76&Yes\\

0~dB&231.42~px&152.87~px&16.12&Yes\\

\hline

\end{tabular}
\vspace{-10pt}
\label{tab:denoising-metrics}
\end{center}
\end{table}

\subsection{System evaluation}
YuNet \cite{wu2023yunet} performs face detection here. In the baseline configuration, the face detection module processes each frame completely. For the other configurations, the module scans the ROIs only. 
To test for the worst case scenarios, all frames are processed regardless of the VAD output.

\begin{figure}[!htbp]
    \centering
    \includegraphics[width=\columnwidth]{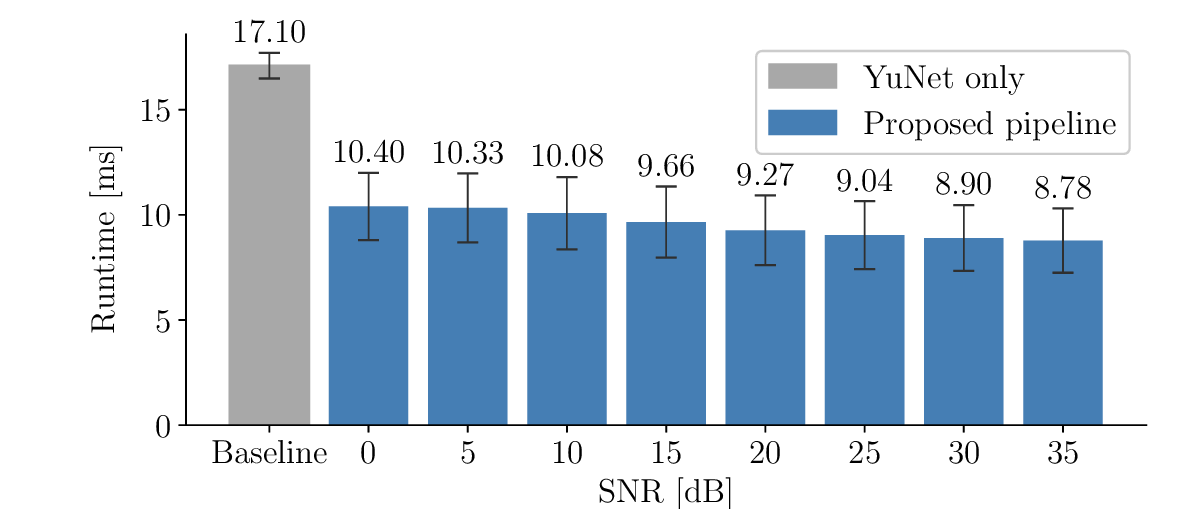}
    \caption{Average runtime of the system to process a single video frame.}
    \label{fig:runtime-plot}
\end{figure}

\begin{figure}[!htbp]
    \centering
    \includegraphics[width=0.95\columnwidth]{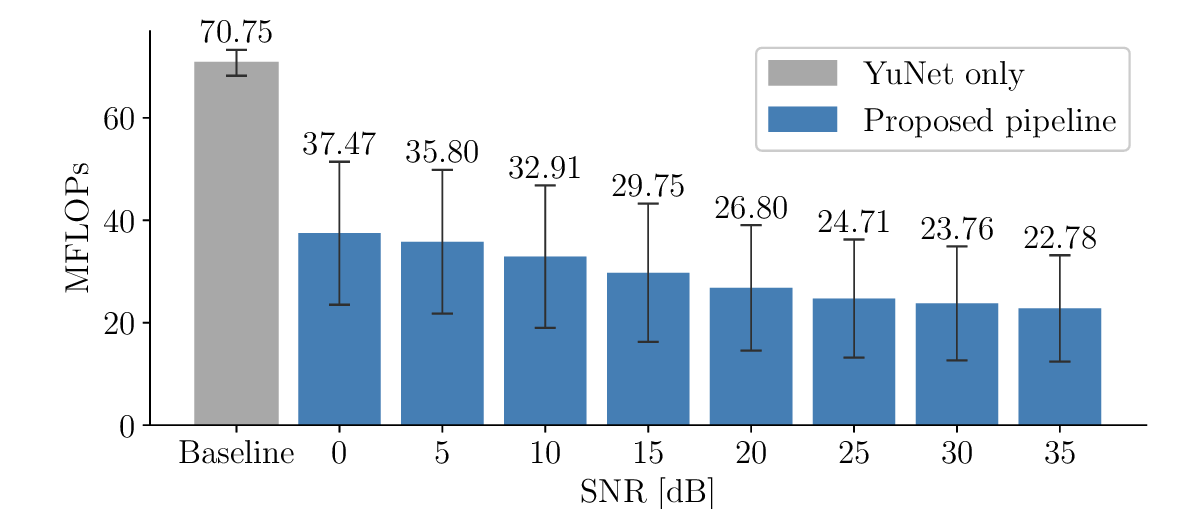}
    \caption{Average number of FLOPs to process a single video frame.}
    \label{fig:mflops-plot}
    \vspace{-10pt}
\end{figure}

To determine how the system compared in terms of computational load, we measure the average runtime and the average number of floating-point operations (FLOPs) to process each video frame in the dataset on an Intel Core i5 and 8GB of RAM. The runtime is measured with \texttt{perf\_counter(\,)\,}from the Python time module while the FLOPs are measured by summing the results of the \texttt{PAPI\_SP\_OPS} and \texttt{PAPI\_DP\_OPS} events of PAPI \footnote{\url{https://icl.utk.edu/papi/}}. 
The estimations by PAPI can vary for the same amount of input data, and the results shown correspond to an average made over numerous runs. 
Figures \ref{fig:runtime-plot} and  \ref{fig:mflops-plot} show how the system scores.

The pipeline was able to reduce the runtime by a factor of 1.95 at a SNR of 35~dB and by a factor of 1.61 at 0~dB which is considerable. These results are further corroborated by the number of floating-point operations that diminishes by a factor of 3.11 at a SNR of 35~dB and by a factor of 1.88 at 0~dB. The variations for the pipeline presented in figure \ref{fig:mflops-plot} are considerable but do not overlap the baseline's. They can be mainly rooted back to the fact that YuNet processes dynamic ROIs. Nevertheless, these results confirm that the proposed pipeline has potential to improve speed.

To determine how relevant the ROIs are, we count the total number of faces that YuNet could detect on full video frames and compare this number to the total count of faces detected on the ROIs. The results are compiled in table \ref{tab:rois-metrics}.

\begin{table}[htbp]
\renewcommand{\arraystretch}{1.3}
\caption{Total number of faces detected in the dataset}
\begin{center}
\begin{tabular}{|c|c|c|c|}
\hline

SNR & Baseline & \multicolumn{2}{|c|}{Proposed pipeline}\\
\hline
\hline

35 dB&72~632&66~424&91.45 \%\\

30 dB&72~632&66~095&91.00 \%\\

25 dB&72~632&65~542&90.24 \%\\

20 dB&72~632&64~544&88.86 \%\\

15 dB&72~632&62~915&86.62 \%\\

10 dB&72~632&59~464&81.87 \%\\

5 dB&72~632&56~626&77.96 \%\\

0 dB&72~632&51~630&71.08 \%\\
\hline

\end{tabular}
\vspace{-5pt}
\label{tab:rois-metrics}
\end{center}
\end{table}

In high SNR, the pipeline is able to capture more than 90\% of the faces initially detected. The method is sensible to the SNR and the results drop to 71.08\% for a SNR of 0~dB. These numbers show that the proposed pipeline offers an interesting trade-off between speed and accuracy and is pertinent for a range of applications.

\section{CONCLUSION}
In this paper, we proposed a unique pipeline to generate voice-targeted ROI and efficiently detect the face of a user in the case of human-robot interactions.
As future work, it would be relevant to 1) train the models on more dynamic noise sources; 2) explore low-dimensional audio input features to reduce the number of parameters in the neural networks; and 3) expand the method to cover the case of multiple speakers at the same time.


\bibliographystyle{IEEEtran}
\bibliography{IEEEabrv,references}

\end{document}